\documentclass[conference]{IEEEtran}
\IEEEoverridecommandlockouts

\usepackage[letterpaper, left=1in, right=1in, bottom=1in, top=0.75in]{geometry}

\usepackage{cite}
\usepackage{amsmath,amssymb,amsfonts}
\usepackage{algorithmic}
\usepackage{graphicx}
\usepackage{subfigure}
\usepackage{textcomp}
\usepackage{acronym}
\usepackage{multirow}

\usepackage{prettyref}
\newrefformat{fig}{Fig.~\ref{#1}}
\newrefformat{conj}{Conjecture~\ref{#1}}
\newrefformat{tab}{Table~\ref{#1}}
\newrefformat{sec}{Section~\ref{#1}}
\newrefformat{subsec}{Section~\ref{#1}}
\newrefformat{subsubsec}{Section~\ref{#1}}
\usepackage[super]{nth}
\usepackage[binary-units]{siunitx}
\usepackage{amsthm}
\usepackage{bm}

\usepackage[right=1in]{geometry}
 
\theoremstyle{definition}

\theoremstyle{remark}

\begin{document}

\title{Fairness-Aware Scheduling in\\
Multi-Numerology Based 5G New Radio
\thanks{The extended and updated version of this work was published in EURASIP Journal on Wireless Communications and Networking \cite{yazar2019a}. https://doi.org/10.1186/s13638-019-1435-z}}

\author{
\IEEEauthorblockN{Ahmet Yazar\IEEEauthorrefmark{1} and H\"{u}seyin Arslan\IEEEauthorrefmark{1}\IEEEauthorrefmark{2}}
\IEEEauthorblockA{\IEEEauthorrefmark{1}Department of Electrical and Electronics Engineering, Istanbul Medipol University, Istanbul, 34810 Turkey\\
}
\IEEEauthorblockA{\IEEEauthorrefmark{2}Department of Electrical Engineering, University of South Florida, Tampa, FL 33620 USA\\
Email: \{ayazar,huseyinarslan\}@medipol.edu.tr}
}

\maketitle

\begin{abstract}
Multi-numerology waveform based 5G New Radio (NR) systems offer great flexibility for different requirements of users and services. Providing \textit{fairness} between users is not an easy task due to inter-numerology interference (INI) between multiple numerologies. This paper proposes two novel scheduling algorithms to provide fairness for all users, especially at the edges of numerologies. Signal-to-noise ratio (SIR) results for multi-numerology systems are obtained through computer simulations.
\end{abstract}

\acresetall


\IEEEpeerreviewmaketitle


\section{Introduction}
\label{sec:introduction}

One of the most remarkable characteristics of New Radio (NR) is its flexibility \cite{ericsson_2016, yazar2018aflexibility, ankarali2017flexible}. The flexibility is needed for application diversity \cite{intel2018}. Requirements of user equipment (UE) and different application groups that include enhanced mobile broadband (eMBB), ultra reliable and low latency communications (URLLC), and massive machine type communications (mMTC) can only be met with a flexible wireless system \cite{seda2018}. One of the most important parts of this flexibility for 5G is originated by employing a multi-numerology waveform design \cite{globecom2018}. Multi-numerology systems provide suitable parameters for different types of services at a time.

A disadvantage of the multi-numerology systems is the inter-numerology interference (INI) that is a leakage between different numerologies, resulting in many challenges and research opportunities \cite{globecom2018}. INI problem shows some similarities with cell edge and fairness problems in Long-Term Evolution (LTE). Non-overlapped two numerologies in frequency have common points with a UE at the edges of two cells in LTE. INI is more effective at the edges of different numerologies and signal-to-interference ratio (SIR) of the edge UEs is low as a result. It causes unfairness for the edge UEs of multiple numerologies like a UE at the edges of two cells.

In this paper, fairness of the UEs on the numerology edges increased by minimizing the effects INI while maintaining the spectral efficiency with fixed guard bands. The overall fairness is also enhanced by our scheduling methods. The proposed fairness-aware algorithms are easily implementable with 3GPP standard.

The rest of the paper is organized as follows: Section~\ref{sec:flexibility} presents some preliminaries and assumptions on multi-numerology systems. Our novel scheduling algorithms are described in Section~\ref{sec:opportunity}. In Section~\ref{sec:simulation}, simulation results for the proposed algorithms are explained. Finally, Section~\ref{sec:conclusion} gives the conclusion.


\section{Assumptions}
\label{sec:flexibility}

\begin{figure}[t]
  \centering
  \includegraphics[width=7cm]{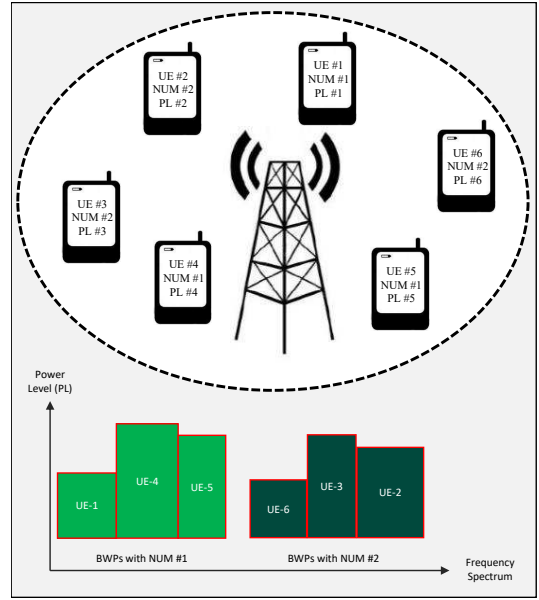}
  \caption{An example resource allocation in frequency spectrum for the multiple numerologies (NUM) with different user power levels.}
\label{fig:Fig1}
\end{figure}

Table~\ref{tab:numerology} shows the numerology parameters including subcarrier spacing ($\Delta{f}$), CP duration ($T_{CP}$), and slot duration for data channels in NR according to 3GPP standard documents \cite{globecom2018} and \cite{3gpp.38.211}. It is assumed that UEs are synchronous to each other. We allocate UEs or bandwidth parts (BWP) with same numerologies contiguously in the frequency domain \cite{intel2018, ericsson2017, globecom2018}. It is also assumed that UEs are non-overlapping to each other and each numerology block that consists of multiple carriers is shared by multiple UEs. Orthogonal Frequency Division Multiplexing (OFDM) is employed and each UEs have different power levels. Fig.~\ref{fig:Fig1} shows the basic scenario. We assumed that numerology selection procedures have completed regarding different user and service requirements \cite{yazar2018aflexibility}. For example, base station (BS) assigns NUM-1 to UE-1, 4, 5; and NUM-2 to UE-6, 3, 2 in Fig.~\ref{fig:Fig1}.

\begin{table}
\renewcommand{\arraystretch}{1.3}
\caption{Numerology Structures for Data Channels in NR}
\label{tab:numerology}
\centering
\begin{tabular}{|c|c|c|c|}
\hline
$\Delta{f}$ & $T_{CP}$ & Slot & $\#$ of Symbols \\
$(\si{\kilo\hertz})$ & ($\si{\micro\second}$) & Duration (\si{\milli\second}) & in One Slot \\
\hline
15 & 4.76 & 1 & 14 \\
\hline
30 & 2.38 & 0.5 & 14\\
\hline
60 & 1.19 $|$ 4.17 & 0.25 & 12 $|$ 14 \\
\hline
120 & 0.60 & 0.125 & 14 \\
\hline
\end{tabular}
\end{table}


\section{Fairness-Aware Scheduling in Multi-Numerology Systems}
\label{sec:opportunity}

In this section, power difference problem for the edge users of numerologies is analyzed. Then, novel algorithms are proposed to increase fairness by scheduling users at the edges of numerologies more carefully.

\subsection{Power Difference for the Edge Users of Different Numerologies}

INI can be defined as the leakage from one numerology to the other due to the large side lobes of waveform blocks \cite{globecom2018}. Various estimation models and cancellation methods for INI are presented in \cite{tafazolli_subband, pekoz2017adaptive, zhang2018}. These types of interferences are generally concentrated at the edge subcarriers of numerologies because of the large side lobes \cite{demir2017theimpact}. Besides, there is a guard requirement between different numerologies \cite{yoshihisa_2017_1, ericsson_2016, demir2017theimpact}.

In addition to the INI problem for the UEs on numerology edges, power difference is another issue for multi-numerology systems \cite{demir2017theimpact}. SIR degradation occurs especially at the edge UEs in different numerologies. Power offset affects SIR negatively. SIR distribution goes down from center to edge users. Hence, fairness for the edge UEs of numerologies needs to be provided while maintaining the other performance criteria.

Power offset of the edge UEs can be minimized to increase fairness for the edge UEs. Also, minimizing a variance between SIR values for different cases aims the same motivation. SIR values of one UE should not change noticeably with time. Weak UEs are affected easily by high power offsets like in the near-far problem for a cell. It causes higher SIR variances for these UEs. There is a need to balance SIR to preserve the fairness of users and protect weak users.

A lower power offset can also be useful to minimize guard necessities between different numerologies under desired SIR \cite{demir2017theimpact}. In that case, spectral efficiency can be increased due to the fewer guards. However, we increase fairness and SIR for the weak UEs to protect them under fixed guards and spectral efficiency. Fairness requirement has a higher priority in our scenario.

Non-edge inner UEs of different numerologies are affected by INI and power offset less than UEs at the edges. However, it is also possible to increase SIR for inner UEs in addition to the fairness of edge UEs. Putting users with low power levels at the edges of numerologies enables enhancement on SIR of inner UEs. This operation can be applied together with the low power offset solution. Therefore, edge UE of one numerology will not affect the inner UEs of the other numerology too much. In other words, there are two goal functions. The first of them is about the interaction between edge UEs of the numerologies and it is more important because most of the INI is concentrated on the numerology edges. We need to maximize SIR at the edge users. The second goal function is focused on the interaction between one edge UE of one numerology and the inner UEs of the other numerology. In this case, we can also enhance SIR on the inner UEs.

In our system, it is assumed that there are multiple users with different power levels in the same numerology. However, all users have different power levels and numerology parameters in \cite{demir2017theimpact}. Authors of \cite{demir2017theimpact} put each user in a specific place regarding their power levels. It causes a low frequency dependent scheduling flexibility. They aim to minimize guard necessities with a fixed SIR and fairness in their scheduling algorithm. We propose scheduling algorithms that focus only edge users of the numerologies to maximize fairness for edge and inner UEs of contiguous multiple numerologies.

\subsection{Proposed Algorithms for Fairness-Aware Scheduling}

The proposed fairness-aware scheduling algorithms are presented in Fig.~\ref{fig:Fig2}. The upper demonstration shows a random scheduling case, and the below ones show the proposed scheduling mechanisms. The first algorithm maximizes the fairness of edge UEs. However, the second algorithm also enhances the SIR of inner UEs by sacrificing a little bit fairness from the edge UEs. There is a small trade-off between these two algorithms.

\subsubsection{Algorithm 1: Scheduling Based on Edge-User Fairness}

\begin{figure}[t]
  \centering
  \includegraphics[width=7cm]{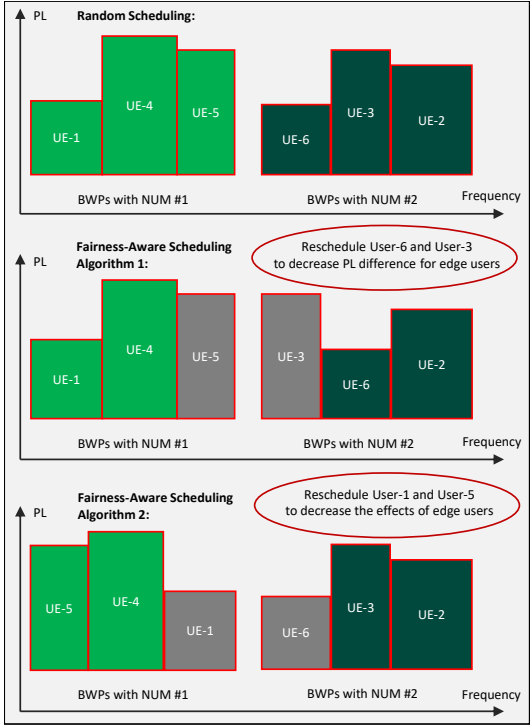}
  \caption{Fairness-aware scheduling in multi-numerology systems.}
\label{fig:Fig2}
\end{figure}

The proposed method schedules UEs as a function of power offsets between the UEs for different numerologies. In Fig.~\ref{fig:Fig2}, frequency positions of UE-6 and UE-3 are replaced with each other in the same numerology. Hence, the power offset between edge UEs (UE-5 and UE-3) are minimized to ensure that SIR is maximized at the edges of numerologies.

There can be more than two numerologies at a time but our algorithm works based on numerology pairs like in Fig.~\ref{fig:Fig2}. The algorithm needs to be employed for each of the contiguous two numerologies. For this reason, it is assumed that there are two numerologies in the remaining parts of the paper.

There are $D$ users ($u_{1,1}$, $u_{1,2}$, ... ,$u_{1,D}$) for NUM-1, and $E$ users ($u_{2,1}$, $u_{2,2}$, ... ,$u_{2,E}$) for NUM-2. Power levels of these users are ($p_{1,1}$, $p_{1,2}$, ... ,$p_{1,D}$) and ($p_{2,1}$, $p_{2,2}$, ... ,$p_{2,E}$), respectively. Then, there are totally $DxE$ possibilities for the power offsets, $PO$, between UE pairs with different numerologies. The smallest power difference selection is made using Eq.~(\ref{eq:eq1}) and Eq.~(\ref{eq:eq2}). Then, the resulting UE pair, ($s,t$), can be located at the edges of numerologies to increase fairness for edge UEs.

\begin{equation}
PO(s,t) = |p_{1,s}-p_{2,t}|
\label{eq:eq1}
\end{equation}

\begin{equation}
{(s,t)}^{*} = \mathop {\mathrm {argmin}} _{(s,t)} PO(s,t)
\label{eq:eq2}
\end{equation}
where $s$ and $t$ are UEs for NUM-1 and NUM-2, respectively.

\begin{figure*}[t]
  \centering
  \subfigure[Case 1: All users have equal power levels.]{\includegraphics[width=7cm]{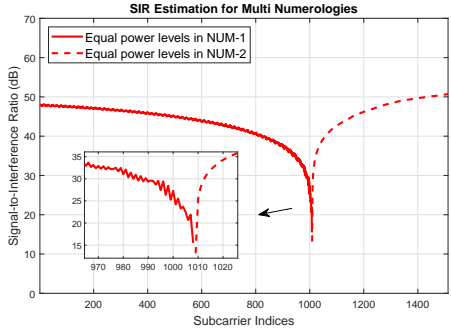}}\qquad
  \subfigure[Case 2: Edge user of NUM-2 has higher power level than the other users in NUM-1 and NUM-2.]{\includegraphics[width=7cm]{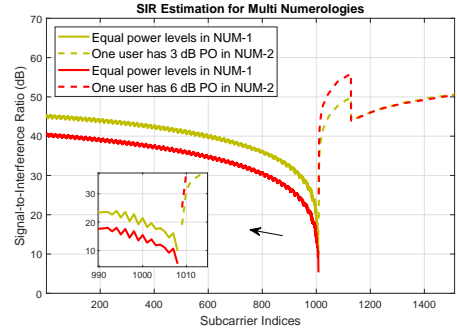}}\\
  \subfigure[Case 3: One inner user of NUM-2 has higher power level than the other users in NUM-1 and NUM-2.]{\includegraphics[width=7cm]{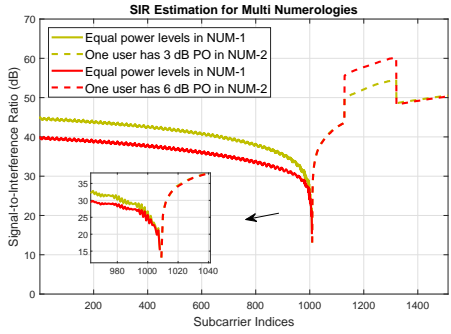}}\qquad
  \subfigure[Case 4: Edge users of NUM-1 and NUM-2 have higher power levels than the inner users in NUM-1 and NUM-2. There is not any power offsets between the edge users.]{\includegraphics[width=7cm]{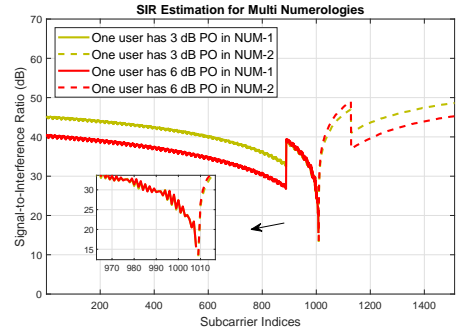}}
  \caption{Performance analysis results for different cases. NUM-1 has narrow subcarriers with 15 kHz and NUM-2 has wide subcarriers with 30 kHz. There is not any guard bands between numerologies. The number of usable subcarriers for edge users in each numerology is 120.}
\label{fig:Fig3}
\end{figure*}

\subsubsection{Algorithm 2: Scheduling Based on Overall Fairness}

The second method schedules UEs as a function of 1) power offsets between the UEs for different numerologies, and 2) power levels of all UEs. If we want to increase overall fairness in parallel to edge user fairness as mentioned in the previous subsection, there are some other steps. As an example, frequency positions of UE-1 and UE-5 are replaced with each other in Fig.~\ref{fig:Fig2} according to a decision of this algorithm. Hence, the power offset between edge UEs (UE-1 and UE-6) are selected small and also power levels of the edge users are selected low to ensure that overall fairness is enhanced.

In the second algorithm, a vector of edge candidate pairs, $\bm{H}$, is formed using a threshold given in Eq.~(\ref{eq:eq3}).

\begin{equation}
TH_p = r \times PO(s,t)^{*}
\label{eq:eq3}
\end{equation}
where $r$ is the threshold factor and $r\geq1$. $TH_p$ is the maximum threshold for the power offsets between UE pairs. If $r$ increases, the more UE pairs will be candidates for the numerology edges. It is possible to adjust it for different scenarios. If it is 1, then the only candidate will be the UE pair of $(s,t)^{*}$. UE pairs that have less power offsets than $TH_p$ are added into the vector of edge candidate pairs, $\bm{H}$.

In the next step, two power levels ($p_{1,s}$ and $p_{2,t}$) in each UE pair of $\bm{H}$ is averaged and $\bm{PL}$ vector is formed using the averaged power level values. The lowest value of $\bm{PL}$ can be employed to decide on UE pair of $(s,t)^{*}$ using Eq.~(\ref{eq:eq4}) to schedule them as the edge UEs of multiple numerologies.

\begin{equation}
{(s,t)}^{*} = \mathop {\mathrm {argmin}} _{(s,t)} \bm{PL}
\label{eq:eq4}
\end{equation}

We try to focus on mostly two users at the edges in our proposed algorithms. Then, frequency dependent scheduling flexibility does not lose. Additionally, computational complexity of the algorithms are very low since they are simple methods.


\section{Simulation Results}
\label{sec:simulation}

In the simulations, it is assumed that there are three UEs in each numerology like in Fig.~\ref{fig:Fig2}. Some other simulation parameters are provided in Table~\ref{tab:simulation}.

$\SI[parse-numbers = false]{\Delta f_\text{ref}}{\kilo\hertz}$
and $\SI[parse-numbers = false]{2^k\times\Delta{f}_{\text{ref}}}{\kilo\hertz}$ subcarrier spacings are used for two numerologies, where $2^k$ is the scaling factor and $k$ is a positive integer. $N_{ref}$-point and $N_{ref}/(2^k)$-point inverse fast Fourier transform (IFFT) blocks are employed by NUM-1 and NUM-2, respectively. After each IFFT operation, CP samples are added with a ratio of $CP_R$ to every OFDM symbol in each numerology. Wireless channel and noise are ignored to just focus on INI in the simulation results. At the receiver side, $N_{ref}$-point and $N_{ref}/(2^k)$-point fast Fourier transform (FFT) blocks are used by NUM-1 and NUM-2, respectively. The same structure is used in the rest of this section.

\begin{table}[!b]
\renewcommand{\arraystretch}{1.3}
\caption{Simulation Parameters}
\label{tab:simulation}
\centering
\begin{tabular}{|l|c|c|}
\hline
The number of users for NUM \#1 & $D$ & 3\\
\hline
The number of users for NUM \#2 & $E$ & 3\\
\hline
Reference value for $\Delta{f}$ & $\Delta{f}_{ref}$ & 15 kHz \\
\hline
The scaling factor for $\Delta{f}$ & $k$ & 1 \\
\hline
Reference size of IFFT/FFT blocks & $N_{ref}$ & 4096 \\
\hline
CP Ratio & $CP_R$ & 1/16 \\
\hline
Threshold factor & $r$ & 2 \\
\hline
\end{tabular}
\end{table}

\subsection{Performance Analysis}

Power offsets of the UEs alternate between 0 dB, 3 dB, and 6 dB. INI and SIR estimations are done for each of the used subcarriers separately. Monte Carlo method is applied to increase the statistics in performance results. The number of tests is 1000 and different set of random data is used in each of these tests. Thereafter, the average INI and SIR on the subcarriers are estimated. SIR results are presented in Fig.~\ref{fig:Fig3} with the below comments:

\begin{enumerate}
  \item In Fig.~\ref{fig:Fig3}(a) (Case-1), all users have equal power levels. Average SIR results for two UEs in the same numerology differ at least 7 dB. It directly shows that edge UEs face with an unfairness.
  \item In Fig.~\ref{fig:Fig3}(b) (Case-2), increasing the power level of the NUM-2 edge UE 3 dB results with 5.7 dB SIR decrement in the NUM-1 edge UE. PO increment also affects the NUM-1 inner UEs with 4.8 dB. However, there is more than 10 dB SIR difference between the edge UE and inner UEs for NUM-1.
  \item In Fig.~\ref{fig:Fig3}(c) (Case-3), increasing the power level of the NUM-2 inner UE 3 dB results with 2.8 dB SIR decrement in the NUM-1 edge UE. Therefore, PO increment for the inner UEs do not affect edge UEs too much compared to Case-2.
  \item In Fig.~\ref{fig:Fig3}(d) (Case-4), increasing the power level of the NUM-1 and NUM-2 UEs symmetrically results with a small SIR increment in the edge UEs of two numerologies. However, inner UEs are affected by the power levels of edge UEs in proportion.
\end{enumerate}

\subsection{Simulation Results of the Algorithms}

In this subsection, power level offsets are generated 1000 times randomly between 0 dB and 10 dB. Proposed scheduling algorithms compared with a random scheduling case. The main aim of our scheduling algorithms is that minimizing the variance between SIR values for different cases. SIR values of one user should not change noticeably with time. There is too much fluctuation in SIR of edge users of numerologies for the random scheduling scenario. Proposed algorithms balance SIR to preserve the fairness between users.

\begin{figure}[t]
  \centering
  {\includegraphics[width=5.8cm]{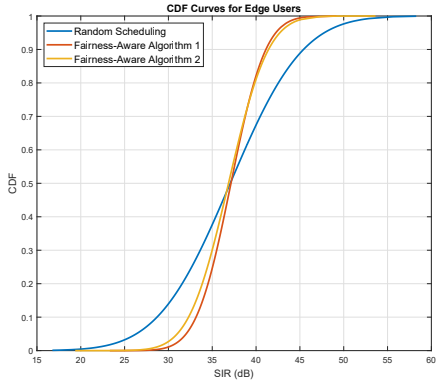}}
  \caption{The proposed fairness-aware scheduling algorithms compared to random scheduling. Only edge UEs are included. NUM-1 has narrow subcarriers with 15 kHz and NUM-2 has wide subcarriers with 30 kHz.}
\label{fig:Fig4}
\end{figure}

Cumulative distribution function (CDF) curves for the edge UEs are presented in Fig.~\ref{fig:Fig4}. Here, the number of usable subcarriers are taken equally for all users. CDF curves show that the variance in SIR for our algorithms are lower than the random scheduling case. Therefore, fairness of the edge UEs is enhanced by using fairness-aware scheduling methods.

\begin{figure}[t]
  \centering
  \subfigure[There are 168 and 84 usable subcarriers for the edge UEs of NUM-1 and NUM-2, respectively.]{\includegraphics[width=5.9cm]{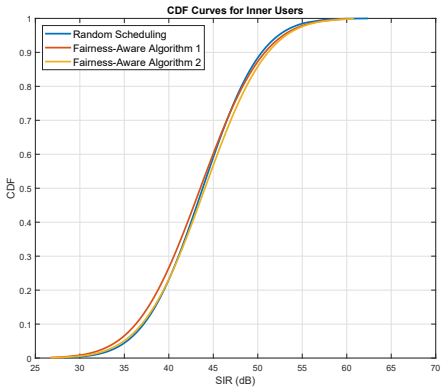}}\qquad
  \subfigure[There are 672 and 336 usable subcarriers for the edge UEs of NUM-1 and NUM-2, respectively.]{\includegraphics[width=5.9cm]{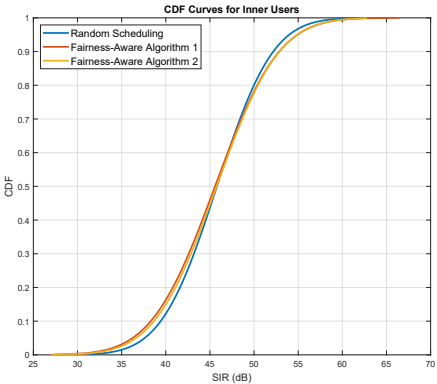}}\\
  \caption{CDF curves for the inner users. NUM-1 has narrow subcarriers with 15 kHz and NUM-2 has wide subcarriers with 30 kHz.}
\label{fig:Fig5}
\end{figure}

As it can be seen from Fig.~\ref{fig:Fig5}, Algorithm 2 increase fairness of inner UEs slightly compared to Algorithm 1 in return for a small loss in the fairness of edge UEs. Additionally, the difference between Fig.~\ref{fig:Fig5}(a) and Fig.~\ref{fig:Fig5}(b) is the number of used subcarriers at the edge UEs of numerologies. There are four times more subcarriers at the edge UEs of two numerologies in Fig.~\ref{fig:Fig5}(b). When the number of subcarriers of edge UE of NUM-1 increase, INI effects of NUM-2 edge UE at inner UEs of NUM-1 show decrease due to decaying side lobes of NUM-2. Therefore, Algorithm 2 behaves like Algorithm 1 if the edge UEs have wide bandwidths.


\section{Conclusion}
\label{sec:conclusion}

5G systems are designed to achieve better flexibility in an effort to support diverse services and user requirements. It is possible to apply our adaptive scheduling algorithms in multi-numerology 5G systems to increase fairness between UEs. The proposed algorithms can be combined with numerology selection methods and adaptive guard concept. Meanwhile, fairness needs to be handled cautiously. This type of adaptive resource allocation algorithms needs to be designed and optimized for NR. Implementation dependent parts of the 5G standardization offer many other flexibility aspects that can be exploited as research opportunities.



\begin{thebibliography}{1}

\bibitem{yazar2019a}
A. Yazar and H. Arslan, ``Reliability Enhancement in Multi-Numerology-Based 5G New Radio Using INI-Aware Scheduling,'' EURASIP Journal on Wireless Communications and Networking, vol. 2019, no. 110, pp. 1-14, May 2019.

\bibitem{ericsson_2016}
A. A. Zaidi et al., ``Waveform and numerology to support 5G services and requirements,'' IEEE Communications Magazine, vol. 54, no. 11, pp. 90-98, Nov. 2016.

\bibitem{yazar2018aflexibility}
A. Yazar and H. Arslan, ``A flexibility metric and optimization methods for mixed numerologies in 5G and beyond,'' IEEE Access, vol. 6, no. 1, pp. 3755-3764, Feb. 2018.

\bibitem{ankarali2017flexible}
Z. Ankarali, B. Pekoz, and H. Arslan, ``Flexible Radio Access Beyond 5G: A Future Projection on Waveform, Numerology Frame Design Principles,'' IEEE Access, vol. 5, no. 1, pp. 18295-18309, Dec. 2017.

\bibitem{intel2018}
J. Jeon, ``{NR} wide bandwidth operations,'' IEEE Communications Magazine, vol. 56, no. 3, pp. 42-46, Mar. 2018 

\bibitem{seda2018}
S. Dogan, A. Tusha, and H. Arslan, ``OFDM with index modulation for asynchronous {mMTC} networks,'' Sensors, vol. 18, no. 4, pp. 1-15, Apr. 2018.

\bibitem{globecom2018}
A. Yazar and H. Arslan, ``Flexible Multi-Numerology Systems for 5G New Radio,'' River Publishers Journal of Mobile Multimedia, vol. 14, no. 4, pp. 367-394, Oct. 2018.

\bibitem{3gpp.38.211}
3rd Generation Partnership Project (3GPP), ``NR; Physical channels and modulation,'' Technical Specification 38.211, ver. 15.1.0, Apr. 2018.

\bibitem{ericsson2017}
S. Parkvall et al., ``{NR}: the new {5G} radio access technology,'' IEEE Communications Standards Magazine, vol. 1, no. 4, pp. 24-30, Dec. 2017.

\bibitem{tafazolli_subband}
L. Zhang et al., ``Subband filtered multi-carrier systems for multi-service wireless communications,'', IEEE Transactions on Wireless Communications, vol. 16, no. 3, pp. 1893-1907, Mar. 2017.

\bibitem{pekoz2017adaptive}
B. Pekoz, S. Kose, and H. Arslan, ``Adaptive Windowing of Insufficient CP for Joint Minimization of ISI and ACI Beyond 5G,'' IEEE Int. Symp. Personal, Indoor, and Mobile Radio Commun. (PIMRC), Montreal, QC, Oct. 2017, pp. 1-5.

\bibitem{zhang2018}
X. Zhang et al., ``Mixed Numerologies Interference Analysis and Inter-Numerology Interference Cancellation for Windowed OFDM Systems,'' accepted for publication in IEEE Transactions on Vehicular Technology, 2018.

\bibitem{demir2017theimpact}
A. F. Demir and H. Arslan, ``The Impact of Adaptive Guards for 5G and Beyond,'' IEEE Int. Symp. Personal, Indoor, and Mobile Radio Commun. (PIMRC), Montreal, QC, Oct. 2017, pp. 1-5.

\bibitem{yoshihisa_2017_1}
P. Guan et al., ``5G field trials: OFDM-based waveforms and mixed numerologies,'' IEEE Journal on Selected Areas in Communications, vol. 35, no. 6, pp. 1234-1243, June 2017.

\end{thebibliography}
\end{document}